\documentclass[]{aa}
\usepackage{graphicx,txfonts,amssymb,natbib}
\renewcommand{\d}{\mathrm{d}}
\authorrunning{C. Fedeli and M. Bartelmann}
\titlerunning
  {Effects of early dark energy on strong cluster lensing}

\begin{document}

\title
  {Effects of early dark energy on strong cluster lensing}

\author{Cosimo Fedeli and Matthias Bartelmann
  \institute
    {Zentrum f\"ur Astronomie, ITA, Universit\"at Heidelberg,
     Albert-\"Uberle-Str. 2, 69120 Heidelberg, Germany}}

\date{\emph{Astronomy \& Astrophysics, submitted}}

\abstract{We use the semi-analytic method developed by Fedeli et al.\
  for computing strong-lensing optical depths to study the statistics
  of gravitational arcs in four dark-energy cosmologies. Specifically,
  we focus on models with early dark energy and compare them to more
  conventional models. Merger trees are constructed for the cluster
  population because strong cluster lensing is amplified by factors of
  two to three during mergers. We find that the optical depth for
  gravitational arcs in the early dark-energy models is increased by
  up to a factor of $\sim3$ compared to the other models because of
  the modified dynamics of cluster formation. In particular, the
  probability for gravitational arcs in high-redshift clusters is
  considerably increased, which may offer an explanation for the
  unexpectedly high lensing efficiency of distant clusters.}


\maketitle

\section{Introduction}

While the present dominance of dark energy is well established
\citep{GO03.1,HA03.1,SP03.1,SP06.1,RE04.3,RE04.4,RI04.1,TE04.1}, its
evolution in particular in the early universe is largely
unconstrained. An interesting class of models for dynamical dark
energy is characterised by a low but non-vanishing density of dark
energy at early times \citep{FE98.1,DO01.1,DO01.2,CA03.2,WE04.1}.
Non-linear structure formation has recently been studied for this class
of models by
\cite{BA05.1} in the framework of the spherical collapse
model. Interestingly, it was found that non-linear structures are
expected to form substantially earlier in such early dark-energy
models if they are normalised so as to be compatible with the
large-scale temperature fluctuation amplitude of the cosmic microwave
background. For two specific models with early dark energy, the
population of galaxy clusters is expected to evolve by approximately
an order of magnitude less strongly than in the standard $\Lambda$CDM
model.

Should this come close to reality, a rich population of massive galaxy
clusters would be present at high redshift which is completely
unexpected in $\Lambda$CDM. Similarly, the dynamical activity within
the cluster population due to substantial mergers with sub-halos would
be shifted or extended towards higher redshift.

The problem of the non-linear evolution of cosmic structures in presence
of dark energy has been recently addressed also from a more
general point of view by several authors.
For instance, \cite{MO04.1}, \cite{ZE05.1},
\cite{MA05.2} and \cite{WA06.1} analyse different aspects of this issue
for both constant and time dependent dark energy equation of state parameter,
allowing for dark energy clustering and coupling to dark matter. They outline
very different properties of the virialized objects depending on the behaviour
of the dark energy fluid.
Additionally, \cite{ZE05.1} and \cite{MA05.3}
explore the outcome of this different non-linear evolution on the predicted
number counts for high mass dark matter haloes (galaxy clusters),
finding several significant effects. They discover in particular that the
number counts of massive structures increase if small scale clustering of dark
energy is allowed, while it decrease if the amount of dark matter coupled to
dark energy grows.

One interesting and due to its non-linearity highly sensitive way for
probing the massive end of the cluster population is the strong
lensing effect. Although the issue is still controversial
\citep{BA98.2, ME00.1,ME03.2,WA03.1,DA05.1,LI05.1,HE05.1}, it seems to
be at least difficult within the $\Lambda$CDM model to reproduce the
observed abundance of strong-lensing events in cluster cores, the
so-called gravitational arcs. Arcs in clusters at high redshift
\citep{HA98.1,TH01.1,ZA03.1,GL03.1} are similarly puzzling because
they indicate that even clusters at $z\gtrsim1$ can already be
concentrated and massive enough to be strong gravitational lenses for
a source population that is not too distant from them.

Dynamical activity in galaxy clusters was identified as highly
important for their strong-lensing abilities
\citep{BA95.1,ME03.1,TO04.1, FE06.1}. The enhancement of the
gravitational tidal (shear) field while clusters are merging with
massive halos can transiently, but substantially increase their
strong-lensing cross sections. As much as about half of the total
optical depth for strong cluster lensing may be contributed by merging
clusters. The effect is strong because mergers can turn clusters into
strong lenses that would otherwise fall below the threshold because
they are not massive or compact enough. Major cluster mergers thus
open the huge, exponentially rising reservoir of moderately massive
clusters for strong lensing.

Cosmological models reconciling an appreciable cluster abundance at
high redshift, and thus also a high level of dynamical cluster
activity, with independent cosmological constraints e.g.~from the CMB
are thus particularly interesting in view of strong cluster
lensing. Sufficiently detailed numerical simulations are costly and
beyond scope for a parameter study. Recently, \cite{FE06.1} have
developed a (semi-)analytic method for computing strong-lensing cross
sections for galaxy clusters with and without taking cluster mergers
into account. This method opens the way to systematically test a
variety of cosmological models for their consequences for strong
cluster lensing. We use it in this paper to study the statistics of
strong cluster lensing in two exemplary cosmologies with early dark
energy and compare them to the $\Lambda$CDM model and a model with a
constant equation-of-state parameter $w>-1$. We do not
focus on other aspects related to the dark energy fluid behaviour like small
scale clustering or coupling to dark matter.

Section~2 reviews extended Press-Schechter theory as it will be needed
later, and Sect.~3 summarises the cosmological models used. The
construction of merger trees and the computation of strong-lensing
cross sections are described in Sects.~4 and 5. Section~6 outlines
expectations, Sect.~7 quantifies the results, and Sect.~8 summarises
and concludes the paper.

\section{Press-Schechter theory}

We begin by reviewing the basic features of the excursion-set approach
to the derivation of the Press-Schechter \citep{PR74.1} mass function
and the conditional mass function of virialised dark-matter
halos. Comprehensive treatments can be found in \cite{BO91.1} and
\cite{LA93.1}. The central physical quantity is the primordial
Gaussian density-fluctuation field $\delta(\vec{x})$, filtered on a
scale $R$ corresponding to a mass $M$. The filtered field
$\delta_M(\vec{x})$ remains a Gaussian random field whose variance
$S=S(M)$ is a monotonically decreasing function of mass. If the filter
is a top-hat function in Fourier space, $\delta_M(\vec{x})$ at a fixed
location $\vec{x}$ performs a random walk as a function of $M$, and
thus of scale $R$.

When $\delta_M(\vec{x})$ rises above a critical, redshift-dependent
threshold $\delta_\mathrm{c}(z)$, a halo of mass $M$ is expected to
form at the location $\vec{x}$ at redshift $z$. The threshold
$\delta_\mathrm{c}(z)$ is usually obtained from the spherical collapse
model by linearly extrapolating the initial overdensity to the time
when the collapsing halo reaches virial equilibrium.  The problem of a
random walk with a fixed absorbing barrier is solved in \cite{CH43.1}
and leads, under the previous assumptions, to the mass function
\begin{equation}\label{eqn:mf}
  n(M,z)=\frac{\rho_{\mathrm{m},0}}{M}\,
  \frac{\delta_\mathrm{c}(z)}{\sqrt{2\pi}\,S^{3/2}D_+(z)}
  \left|\frac{\d S}{\d M}\right|\,\exp\left[
    -\frac{\delta_\mathrm{c}(z)^2}{2SD_+(z)^2}
  \right]\;.
\end{equation}
The mass function is defined such as $n(M,z)dM$ is the comoving number
density of structures with mass between $M$ and $M+\d M$ at redshift
$z$, where $\rho_{\mathrm{m},0}$ is the mean matter density at present
time. Unlike common practice, we explicitely introduced the linear
growth factor $D_+(z)$ instead of incorporating it into the critical
overdensity in order to emphasise that the redshift dependence in
$\delta_\mathrm{c}(z)$ is exclusively due to the evolution of the
spherical collapse model with redshift. In an Einstein-de Sitter
universe, $\delta_\mathrm{c}\approx1.686$ is a constant, and it
evolves only gently in a $\Lambda$CDM universe, but it changes
considerably in particular in the cosmological models with early dark
energy. We thus have to account for its redshift evolution.

The formalism sketched so far can easily be extended to construct the
conditional mass function which quantifies the probability for a halo
of a given mass $M_0$ at a given redshift $z$ to have a progenitor of
a lower mass $M_\mathrm{p}$ at a higher redshift $z+\Delta z$. Since
the variance of the density field $\delta_M(z)$ filtered on a scale
corresponding to a given mass $M$ decreases monotonically with $M$,
this is equivalent to the probability that a halo of variance $S(M_0)$
at a given redshift had a higher variance $S(M_\mathrm{p})$ at a
higher redshift. This probability is given by \citep{LA93.1}
\begin{equation}\label{eqn:con}
  K(\Delta S,\Delta\omega)=\frac{1}{\sqrt{2\pi}}\,
  \frac{\Delta\omega}{\Delta S^{3/2}}\,
  \exp\left[-\frac{\Delta\omega^2}{2\Delta S}\right]\;,
\end{equation}
where $\Delta S=S(M_\mathrm{p})-S(M_0)$, and $\Delta\omega$ represents
the redshift step considered,
\begin{equation}
  \Delta\omega=\frac{\delta_\mathrm{c}(z+\Delta z)}{D_+(z+\Delta z)}-
  \frac{\delta_\mathrm{c}(z)}{D_+(z)}\;.
\end{equation}
In other words, Eq.~(\ref{eqn:con}) gives the probability for a
dark-matter halo to undergo a change in variance $\Delta S$ due to
hierarchical accretion in the redshift interval $\Delta z$.

If we want the probability for the halo of mass $M_0$ to have a
progenitor corresponding to a change in variance lower than $\Delta S$
within the same redshift interval, we simply have to integrate the
above equation, obtaining the cumulative probability distribution
\begin{equation}\label{eqn:cum}
  J(\Delta S,\Delta\omega)=\int_0^{\Delta S}
  K(\Delta \zeta, \Delta\omega)\d\Delta\zeta=
  \mbox{erfc}\left(\frac{\Delta\omega}{\sqrt{2\Delta S}}\right)\;,
\end{equation}
where
\begin{equation}
  \mbox{erfc}(x)\equiv\frac{2}{\sqrt{\pi}}\,
  \int_x^{\infty}\mbox{e}^{-t^2}\d t
\end{equation}
is the complementary error function. Equation (\ref{eqn:cum}) is just
the probability for the mass of the progenitor $M_\mathrm{p}$ to be
larger than the mass corresponding to the variance $\Delta S-S(M_0)$.

\section{Dark-energy models}

\begin{table}[t!]
  \caption{Cosmological parameters of the four models used in the
    present work.}
  \label{tab:cos}
  \begin{center}
    \begin{tabular}{l|l|l|l|l}
      & EDE1 & EDE2 & $w_\mathrm{de} = -0.8$ & $\Lambda$CDM\\
      \hline
      \hline
      $\sigma_8$ & 0.82 & 0.78 & 0.80 & 0.84\\
      $h$ & 0.67 & 0.62 & 0.65 & 0.65\\
      $\Omega_{\mathrm{m},0}$ & 0.33 & 0.36 & 0.30 & 0.30\\
      $\Omega_{\mathrm{de},0}$ & 0.67 & 0.64 & 0.70 & 0.70 
    \end{tabular}
  \end{center}
\end{table}

Dark energy generalises Einstein's cosmological constant, replacing it
by a term varying with redshift in Friedmann's equation. Viable model
universes with dark energy must be adapted to comply with present-day
observational data. Many models motivated by elementary-particle
physics introduce dark energy as a scalar field (called cosmon or
quintessence, see for instance \citealt{WE88.1,PE88.1,PE02.2,BR00.2}),
whose pressure and energy density are related by the perfect-fluid
equation of state
\begin{equation}
  P_\mathrm{de}=w_\mathrm{de}\,\rho_\mathrm{de}\,c^2\;.
\end{equation}
The parameter $w_\mathrm{de}$ is typically a function of cosmic time
or redshift. A cosmological constant has $w_\mathrm{de}=-1$ at all
redshifts. Different models give rise to different (often not
analytic) shapes for the function $w_\mathrm{de}(z)$, and there are
various ways to parameterise a chosen quintessence model.

Negative pressure at all times implies that the dark-energy density
parameter $\Omega_\mathrm{de}(z)$ will fall to zero for increasing
redshift. If, however, the equation-of-state parameter $w$ is allowed
to rise above zero, models can be constructed in which
$\Omega_\mathrm{de}(z)$ settles at a small positive value. The
presence of a non-vanishing dark-energy contribution in early epochs
of the cosmic evolution can have many interesting consequences on the
CMB temperature fluctuations, the geometry and the age of the
Universe, and the linear and non-linear aspects of structure
formation. Dark-matter halos on all mass scales may form substantially
earlier \citep[][see also \citealt{DO03.2}]{BA05.1}, potentially
causing large effects on the statistics of strong-lensing events
\citep[cf.][]{BA03.1}.

\cite{WE04.1} proposed a useful characterisation of cosmological
models with early dark energy which makes use of only three
parameters; the present density parameter in the dark energy
$\Omega_{\mathrm{de},0}$, the present equation-of-state parameter
$w_{\mathrm{de},0}$, and an average value for the dark energy
parameter at early (structure formation) times,
\begin{equation}
  \bar{\Omega}_\mathrm{de,sf}\equiv
  -(\ln a_\mathrm{eq})^{-1}\int_{\ln a_\mathrm{eq}}^0
  \Omega_\mathrm{de}(a)\, \d\ln a\,
\end{equation}
where $a_\mathrm{eq}$ is the scale factor at matter-radiation
equality.

For sufficiently low $w_{\mathrm{de},0}$, such phenomenological early
dark-energy models reproduce the accelerated cosmic expansion in the
present-day Universe similar to cosmological-constant models and can
thus be arranged to agree with low-redshift observations. If
$\Omega_\mathrm{de,sf}$ is small enough, they can also reproduce the
CMB temperature fluctuations. We shall investigate here the same two
early dark-energy models as in \cite{BA05.1}, which have
$\bar{\Omega}_\mathrm{de,sf}=0.04$, spectral indices for the
primordial density-fluctuation power spectrum of $n=1.05$ (model EDE1
henceforth) and $n=0.99$ (hereafter model EDE2). For comparison, we
also include a model with a constant equation-of-state parameter
$w=-0.8$ and a conventional $\Lambda$CDM model for reference.

\begin{figure}[t!]
\begin{center}
  \includegraphics[width=1.0\hsize]{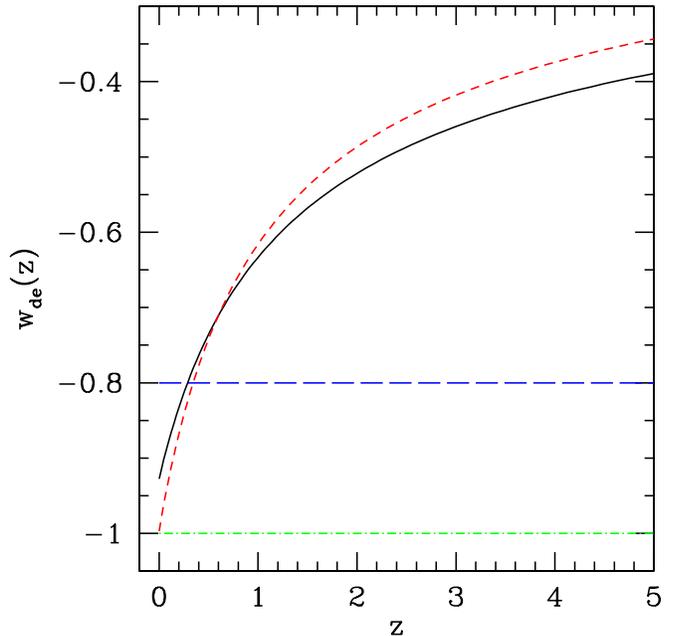}
\end{center}
\caption{The redshift evolution of the equation-of-state parameter for
  the four cosmological models used in this paper. These are a
  $\Lambda$CDM model (green dashed-dotted line), a model with constant
  $w_\mathrm{de}=-0.8$ (blue long-dashed line), and two early
  dark-energy models with different spectral indices for the
  primordial density fluctuations (the black solid line represents
  EDE1 and the red-dashed line EDE2).}
\label{fig:wz}
\end{figure}

The values of the other cosmological parameters, such as the
present-day matter-density parameter $\Omega_{\mathrm{m},0}$, the
dimension-less Hubble constant $h$ and the normalisation of the power
spectrum expressed by $\sigma_8$, were determined such as to match the
power spectrum of the CMB temperature fluctuations
\citep{SP03.1,SP06.1}, constraints from the large-scale structure of
the Universe \citep{TE04.1}, and observations of type-Ia supernovae
\citep{RI04.1}. The values characterising the four cosmological models
used in this paper are listed in Tab.~\ref{tab:cos}, while
Fig.~\ref{fig:wz} shows the redshift evolution of the
equation-of-state parameter in these cosmologies.

As can be noted, the early dark-energy models approach close the
cosmological-constant scenario at very low redshifts.

\section{Merger trees}

We now proceed to use the extended Press-Schechter formalism
summarised in Sect.~2 for a Monte-Carlo realisation of merger trees.
The procedure is quite straightforward, and we refer to \cite{SO99.1}
for a detailed discussion and to \cite{RA02.1} and \cite{CA05.1} for
some applications.

\subsection{Monte-Carlo simulations}

Consider a halo of mass $M_0$ at the present time ($z=0$). If we draw
a random number $r$ in the interval $[0,1]$ and solve the equation
$J(\Delta S,\Delta \omega)-r=0$, we draw a value for the change in the
variance corresponding to the halo compliant with the merger rate
(\ref{eqn:cum}).

Given the variance $S(M_0)$ of the halo's original mass, we obtain a
new value of the variance and convert it to a new mass which is the
mass of the progenitor $M_\mathrm{p}$. If we choose a sufficiently
small time interval, we can assume that the entire change of the
halo's mass is due to a unique, binary merging process with another
halo of mass $\Delta M=M_0-M_\mathrm{p}$. If we repeat this process
for earlier progenitors at subsequent redshift steps, we obtain the
merger history of the original halo up to a given redshift. At the end
of this procedure, we have obtained the value of the halo's mass and
that of its progenitors for each redshift step, i.e.~a merger tree.

The choice of the time interval needs some care. It has to be small to
justify the assumption of binary mergers, but not too small to avoid
that the results be dominated by numerical noise. Following the
rule-of-thumb given by \cite{LA93.1}, we use a time step such that
\begin{equation}\label{eqn:tstep}
  \Delta\omega=\sqrt{\frac{\d S(M_0)}{\d M}\,\Delta M_\mathrm{c}}\;,
\end{equation}
(see also \citealt{SO99.1}) where $\Delta M_\mathrm{c}$ is the mass of
the smallest sub-halo required to be resolved individually. If
$M_\mathrm{p}$ or $\Delta M$ fall below $\Delta M_\mathrm{c}$, the
process does not represent an individual merger, but smooth
accretion. It follows from the above expression that the lower initial
masses $M_0$ require larger time steps.

A set of Monte-Carlo realisations of merger trees is successful if the
population of structures that it produces agrees with the theoretical
mass function at any given redshift. As \cite{SO99.1} pointed out,
this is not strictly so if we consider only binary mergers and smooth
accretion as we are doing here. Several authors \citep{BE05.1} argued
that this may be due to an intrinsic inconsistency in the extended
Press-Schechter formalism, and \cite{SO99.1} suggest that the problem
can be mitigated considering multiple mergers and smooth
accretion. Nonetheless, the difference between the halo-mass
distributions following from the merger-tree simulations and expected
from the mass function is significant only at redshifts beyond our
interest, and we confirmed with several tests the good agreement
between the two halo-mass distributions.

\subsection{Our sample}

We consider a sample of $\mathcal{N}=500$ dark-matter halos whose
present-day masses are \emph{uniformly} distributed within
$M_\mathrm{inf}=10^{14}\,h^{-1}M_\odot$ and
$M_\mathrm{sup}=2.5\times10^{15}\,h^{-1}M_\odot$. It is plausible that
structures with mass below $M_\mathrm{inf}$ at $z=0$ do not contribute
appreciably to the total lensing efficiency (see the discussion in
\citealt{FE06.1}). For each halo, we compute the appropriate time step
from (\ref{eqn:tstep}) and split it into two progenitor halos. Then,
we proceed with the more massive progenitor as the starting point for
the next step. We repeat this procedure until the redshift exceeds the
source redshift $z_\mathrm{s}$ (which is chosen individually for each
halo in the sample, see Sect.~4 for details) or the mass of the halo
falls below $\Delta M_\mathrm{c}$.

\begin{figure}[t!]
\begin{center}
  \includegraphics[width=1.0\hsize]{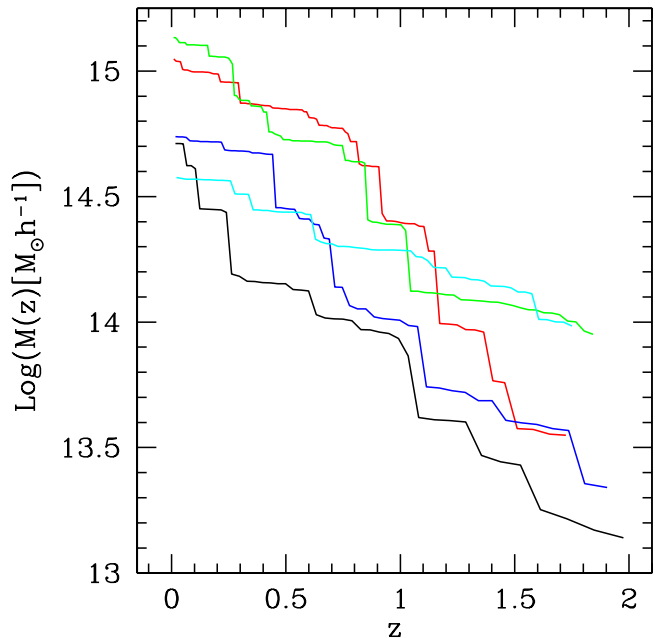}
\end{center}
\caption{Exemplary merger histories for five dark-matter halos
  randomly selected from our sample in a $\Lambda$CDM universe.  The
  merger histories are extended up to the source redshift for each
  individual halo.}
\label{fig:mtree}
\end{figure}

We show in Fig.~\ref{fig:mtree} the merger histories (that is the
evolution of mass with redshift) of five halos selected from our
sample of 500 halos for a $\Lambda$CDM model. Sudden discontinuities
in the mass are evident, each of which corresponds to a merger between
the main halo and a massive sub-halo.

\section{Strong-lensing statistics}

In order to compute the efficiency of dark-matter halos as strong
cluster lenses, specifically for producing long and thin arcs, we
model each halo as a NFW density profile with elliptically distorted
lensing potential. Following \cite{ME03.1}, we adopt an ellipticity
for the iso-potential contours equal to $e=0.3$ for all halos.
Deflection-angle maps for such a lens model can be calculated
analytically \citep{BA96.1}. We then use the fast, semi-analytic
method developed by \cite{FE06.1} to compute the cross sections. We
describe here only its main features and refer the reader to the cited
paper for details.

The lens equation
\begin{equation}\label{eqn:lens}
  \vec{y}(\vec{x})=\vec{x}-\vec{\alpha(\vec{x})}
\end{equation}
relates the (dimensionless) position $\vec y$ of a point source to the
positions $\vec x$ of its images. In the single lens-plane case we
consider here the Jacobian matrix of the mapping (\ref{eqn:lens}) is
symmetric, hence it can be diagonalised through an orthogonal
transformation. The eigenvalues of the Jacobian give the distortions
of the image along the independent directions of the corresponding
eigenvectors. Thus, the ratio of the eigenvalues determines the
length-to-width ratio of the images of point-like sources.

Based upon this consideration, the cross section for gravitational
arcs with length-to-width ratio exceeding a given threshold can be
evaluated as the integral of the inverse of the magnification (the
Jacobian determinant) over the region of the lens plane where the
ratio of the eigenvalues is larger than the threshold. The region of
integration will of course be a stripe surrounding the critical curves
of the mapping. However, this calculation ignores the fact that real
sources are extended and non-circularly shaped, which both enlarge the
cross section compared to point sources.

Thus, to obtain the length-to-width ratio of extended sources, we
convolve the ratio of the eigenvalues with a step function of width
equal to the source size (assumed to be $0.5''$ in radius), and we use
the elegant formalism developed in \cite{KE01.1} to account for source
ellipticities. Source ellipticities are randomly drawn from a flat
distribution between 0.5 and 1.

This method allows us to calculate the cross section
$\sigma_\mathrm{d}$ for arcs with length-to-width ratio exceeding some
given threshold $d$.  We choose $d=7.5$ here and show one plot with
$d=10$ for comparison later.

These semi-analytic cross sections are in excellent agreement with the
results from fully numerical ray-tracing simulations. Moreover, their
computation is substantially faster since the method does not require
costly operations such as finding all images of every source and
refining the source distribution near caustics on an adaptive grid.

\begin{figure}[t]
\begin{center}
  \includegraphics[width=1.0\hsize]{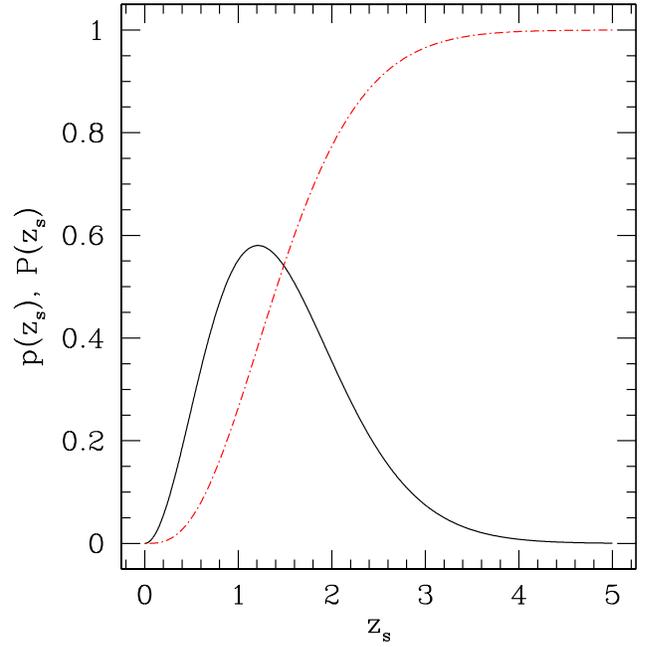}
\end{center}
\caption{The differential (black solid line) and cumulative (red
  dash-dotted line) source-redshift distributions given by
  Eq.~(\ref{eqn:zs}).}
\label{fig:zs}
\end{figure}

We calculate cross sections both ignoring and accounting for merger
processes which transiently increase the lensing efficiency. When a
merger with a sub-halo of mass larger than 5\% of the main halo's mass
occurs, we model the interaction as follows. The two clumps of dark
matter approach each other at a constant speed starting from an
initial distance set to the sum of their virial radii,
$r_{\mathrm{v},1}+r_{\mathrm{v},2}$. The process concludes when the
profiles overlap completely, i.e. when their centres coincide in
projection, and its duration is set to the dynamical timescale
\begin{equation}
  T_\mathrm{dyn}=\sqrt
    {\frac{(r_{\mathrm{v},1}+r_{\mathrm{v},2})^3}{G(M_1+M_2)}}\;.
\end{equation} 
For a fixed and constant source redshift $z_\mathrm{s}$, we can
compute the optical depth $\tau_\mathrm{d}(z_\mathrm{s})$ once the
strong-lensing cross sections for each halo at all redshift steps
between the observer ($z=0$) and the sources ($z=z_\mathrm{s}$) are
known. It is
\begin{equation}\label{eqn:depth}
  \tau_\mathrm{d}(z_\mathrm{s})=\frac{1}{4\pi D_\mathrm{s}^2}\,
  \int_0^{z_\mathrm{s}}\int_0^\infty\,
  N(M,z)\,\sigma_\mathrm{d}(M,z,z_\mathrm{s})\,\d M\d z\;,
\end{equation}
where $D_\mathrm{s}$ is the angular-diameter distance to the source
sphere and $N(M,z)dz$ is the number of structures with mass $M$
included in the shell between redshift $z$ and $z+\d z$. However, in
order to account for the source redshift distribution, we randomly
assigned to each dark-matter halo an individual source redshift
$z_{\mathrm{s},i}$, $i=1,\dots,\mathcal{N}$, drawn from the redshift
distribution of faint blue galaxies given in \cite{SM95.1} (see also
\citealt{BA01.1})
\begin{equation}\label{eqn:zs}
  p(z_\mathrm{s})=\frac{\beta}{z_0^3\Gamma(3/\beta)}\,
  z_\mathrm{s}^2\exp\left[-\left(
    \frac{z_\mathrm{s}}{z_0}
  \right)^\beta\right]\;.
\end{equation}
The parameters $z_0$ and $\beta$ define the average redshift and the
steepness of the high-redshift tail of the distribution,
respectively. In this work, we used the conventional values $z_0=1$
and $\beta=3/2$. Given this choice, the distribution peaks at
$z_\mathrm{m}\simeq1.21$. Figure~\ref{fig:zs} shows this distribution
together with its cumulative function defined by
\begin{equation}
  P(z_\mathrm{s})=\int_0^{z_\mathrm{s}} p(z) \d z\;.
\end{equation}

\noindent
Using the distribution (\ref{eqn:zs}), we can define the average
optical depth as
\begin{eqnarray}\label{eqn:taubar}
  \bar{\tau}_\mathrm{d}&=&\int_0^\infty
  \tau_\mathrm{d}(z_\mathrm{s})\,p(z_\mathrm{s})\d z_\mathrm{s}
  \nonumber\\&=&
  \int_0^\infty\left[
    \int_0^{z_\mathrm{s}}\int_0^\infty
    \sigma_\mathrm{d}(M,z,z_\mathrm{s})\,N(M,z)
    \frac{\d M\d z}{4\pi D_\mathrm{s}^2}
  \right]\,p(z_\mathrm{s})\d z_\mathrm{s}\;.
\end{eqnarray}
Since each halo in our study is characterised by a source redshift
randomly drawn from the distribution (\ref{eqn:zs}), we can omit the
weighting with $p(z_\mathrm{s})$ when we discretise the integral over
source redshift in (\ref{eqn:taubar}). However, this is not possible
for the mass integration, since the masses of the halos are randomly
drawn from a uniform distribution, which requires the weighting with
the halo mass function is necessary.
 
The source-redshift distribution $p(z_s)$ formally extends to an
infinite source redshift, but obviously this is not true in
reality. We set the maximum source redshift to
$z_\mathrm{max}=7.5$. As Fig.~\ref{fig:zs} shows, the probability to
find a source at this redshift can safely be neglected. Since we
operate on a discrete sample of $\mathcal{N}$ halos, each of which is
characterised by a mass $M_i$ and a source redshift
$z_{\mathrm{s},i}$, we can rewrite (\ref{eqn:taubar}) as
\begin{equation}
  \bar{\tau}_\mathrm{d}=\int_0^{z_\mathrm{max}}\left[
    \sum_{i=1}^{\mathcal{N}-1}
    \frac{\sigma_\mathrm{d}(M_i,z,z_{\mathrm{s},i})}
         {4 \pi D^2_{\mathrm{s},i}}
    \int_{M_i}^{M_{i+1}}\,N(M,z)\d M
  \right]\d z\;.
\end{equation}
The integrand of this equation is the optical depth per unit redshift,
i.e.~the contribution to the optical depth from halos at different
redshifts, accounting for the source-redshift distribution
\begin{equation}
  t_\mathrm{d}(z)=\sum_{i=1}^{\mathcal{N}-1}
  \frac{\sigma_\mathrm{d}(M_i,z,z_{\mathrm{s},i})}
       {4 \pi D^2_{\mathrm{s},i}}
  \int_{M_i}^{M_{i+1}}\,N(M,z)\d M\;.
\end{equation}
This will be the central quantity in our strong-lensing analysis.

\section{Expectations}

\begin{figure}[t]
  \includegraphics[width=\hsize]{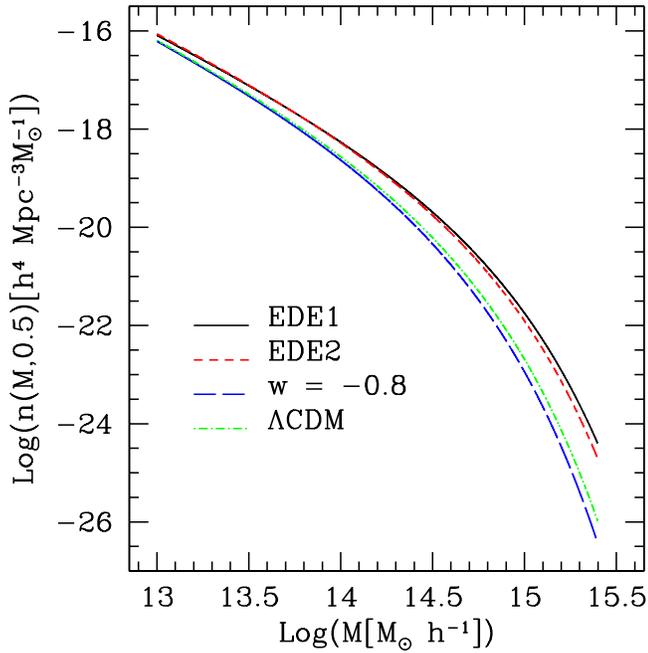}
\caption{The mass function for dark-matter halos in the mass range
  $[10^{13},2.5\times10^{15}]\,h^{-1}M_\odot$ at redshift $z=0.5$ for
  the four cosmological models used in this paper, as labelled in the
  plot.}
\label{fig:mf}
\end{figure}

Before turning to the results, it is useful to evaluate the
expectations in order to gain a better understanding of the
problem. As shown by \cite{BA05.1}, the formation of nonlinear cosmic
structures occupies a larger redshift range in early dark-energy
cosmological models. Structures form earlier and the formation process
lasts longer. This increases the merger probability for a given halo
at high redshift as well as the total number of structures of a given
mass which are found at a given redshift. Figure~\ref{fig:mf} shows
the mass function (\ref{eqn:mf}) at a fixed redshift $z=0.5$ for the
four cosmological models used in this paper.

Evidently, the mass function is lowest for a $\Lambda$CDM model, and
only slightly higher for a model with constant equation-of-state
parameter $w=-0.8$. It is highest (by up to an order of magnitude at
the high-mass tail) for the two early dark-energy models. This
reflects the different halo-formation histories in different
cosmologies. In the EDE1 and EDE2 models, structure formation begins
earlier, hence at a given (suitably low) redshift, the abundance of
halos is larger.

\begin{figure}[t]
  \includegraphics[width=\hsize]{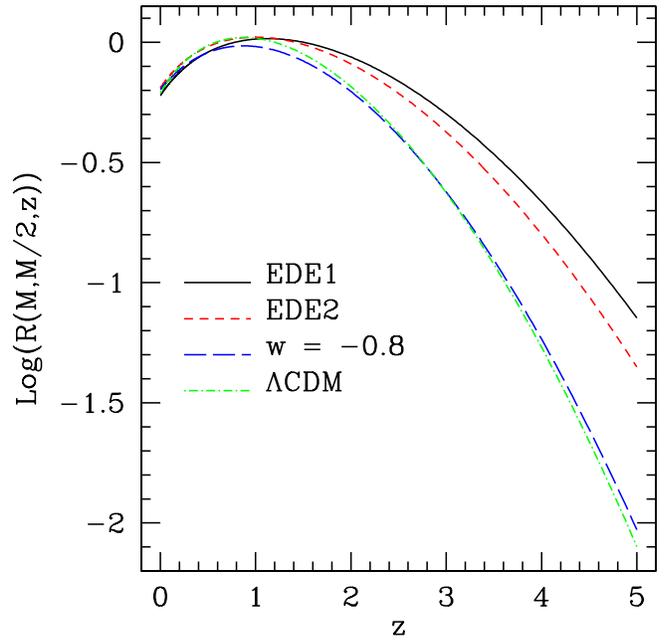}
\caption{The probability for a dark-matter halo of mass
  $M=10^{14}\,h^{-1}M_\odot$ to merge with a sub-halo of mass $M/2$ is
  shown as a function of redshift per unit logarithmic mass of the
  merging sub-halo and per unit logarithmic cosmic time. Results are
  shown for all four cosmological models considered here, as labelled
  in the plot.}
  \label{fig:mr}
\end{figure}

In Fig.~\ref{fig:mr}, we show the merger rate for a halo of mass
$M=10^{14}h^{-1}M_\odot$ and a sub-halo of mass $M/2$ as a function of
redshift. By merger rate, we mean the probability for a halo of mass
$M$ to merge with a sub-halo of mass $M/2$ at redshift $z$ per unit
logarithmic sub-halo mass and per unit logarithmic cosmic time. It can
be obtained as the appropriate limit of the conditional probability
distribution Eq.~(\ref{eqn:con}).

Regarding the merger rate, we also note the difference between the
behaviour of early dark-energy models and of models with constant
equation-of-state parameter. At high redshift, the early dark-energy
merger rate is significantly higher than for the other two models, but
becomes essentially the same below redshift $\sim1.2$.

\begin{figure*}[ht!]
  \centerline
   {\includegraphics[width=0.3\hsize]{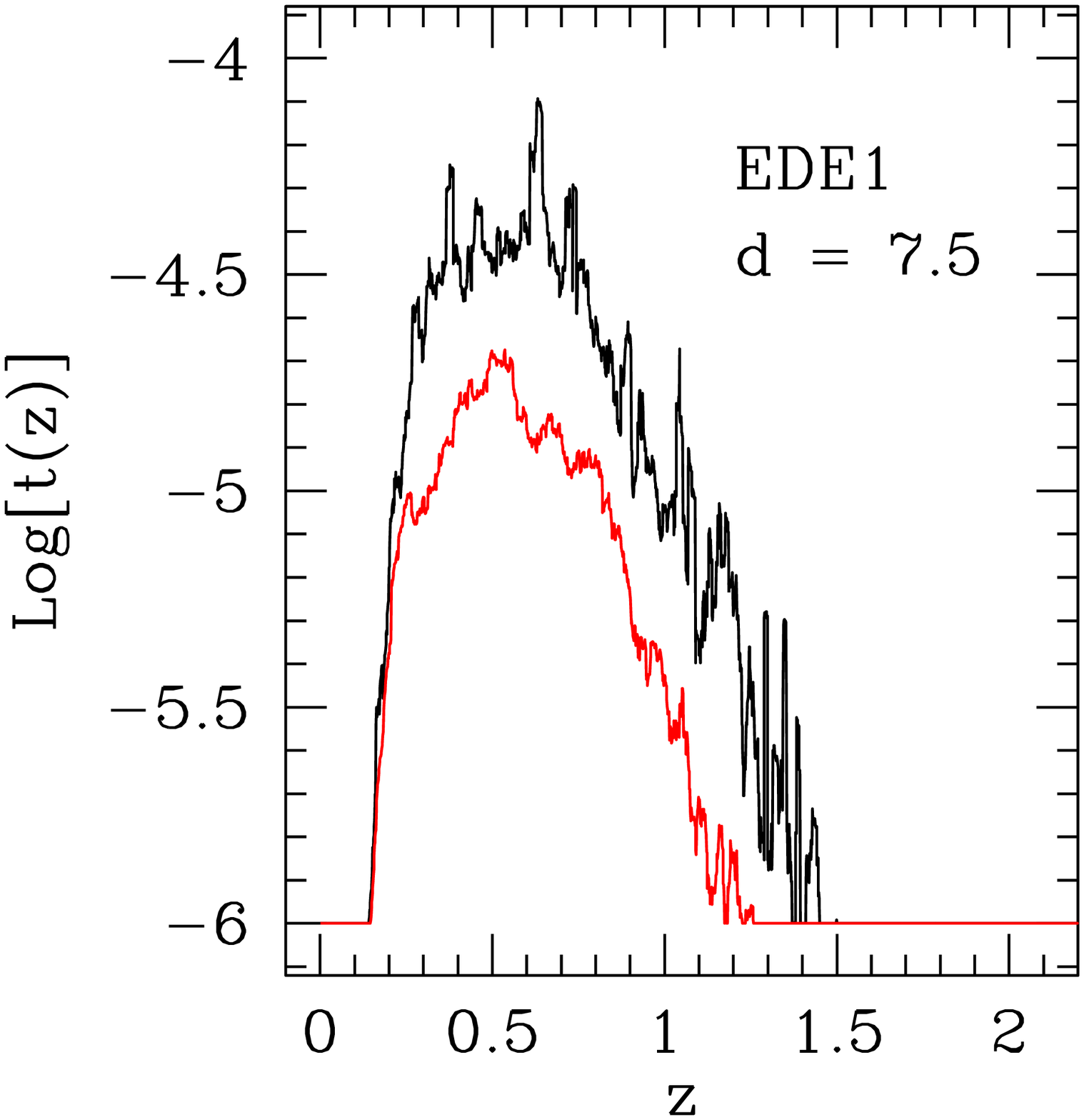}
    \includegraphics[width=0.3\hsize]{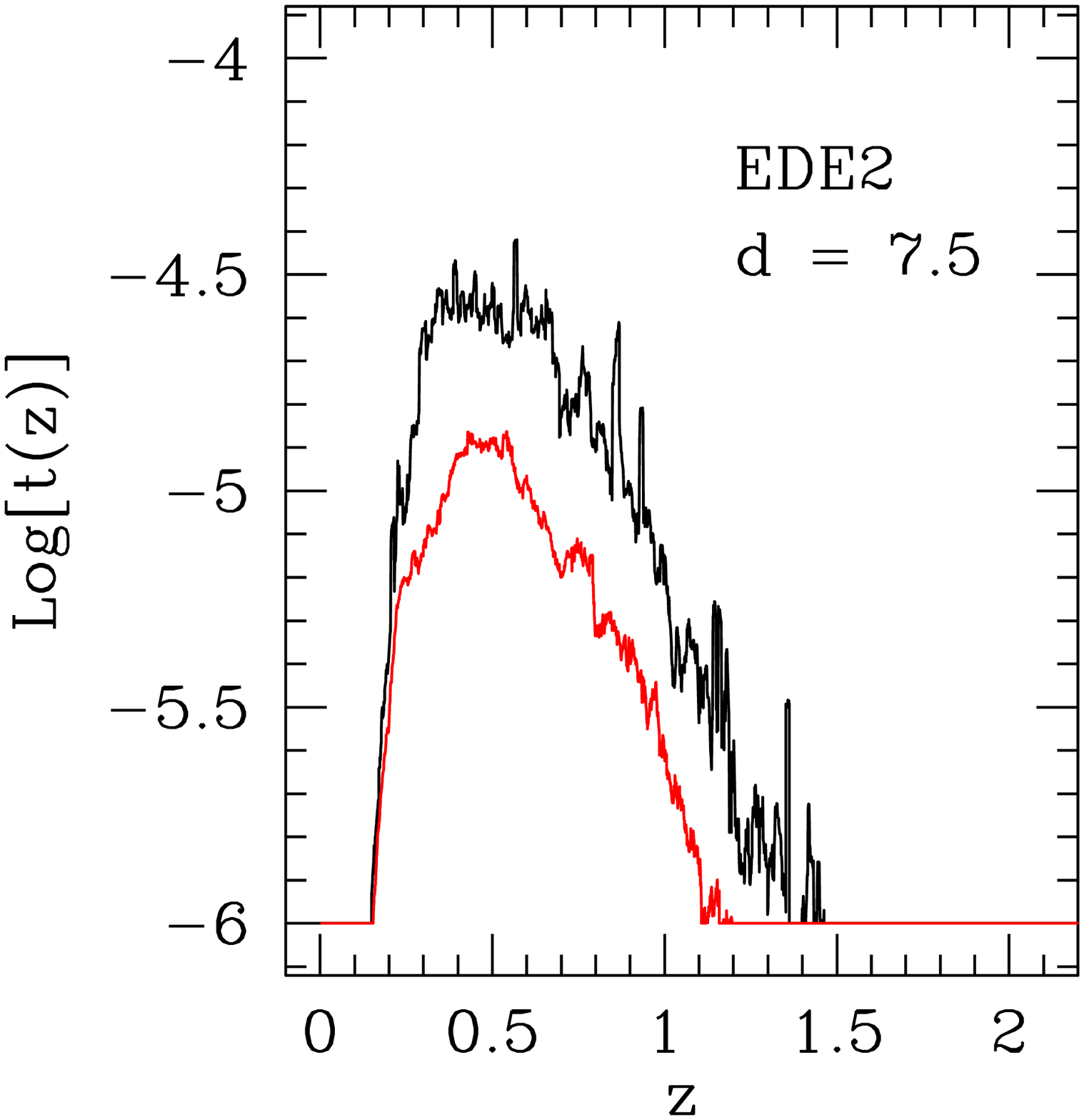}
    \includegraphics[width=0.3\hsize]{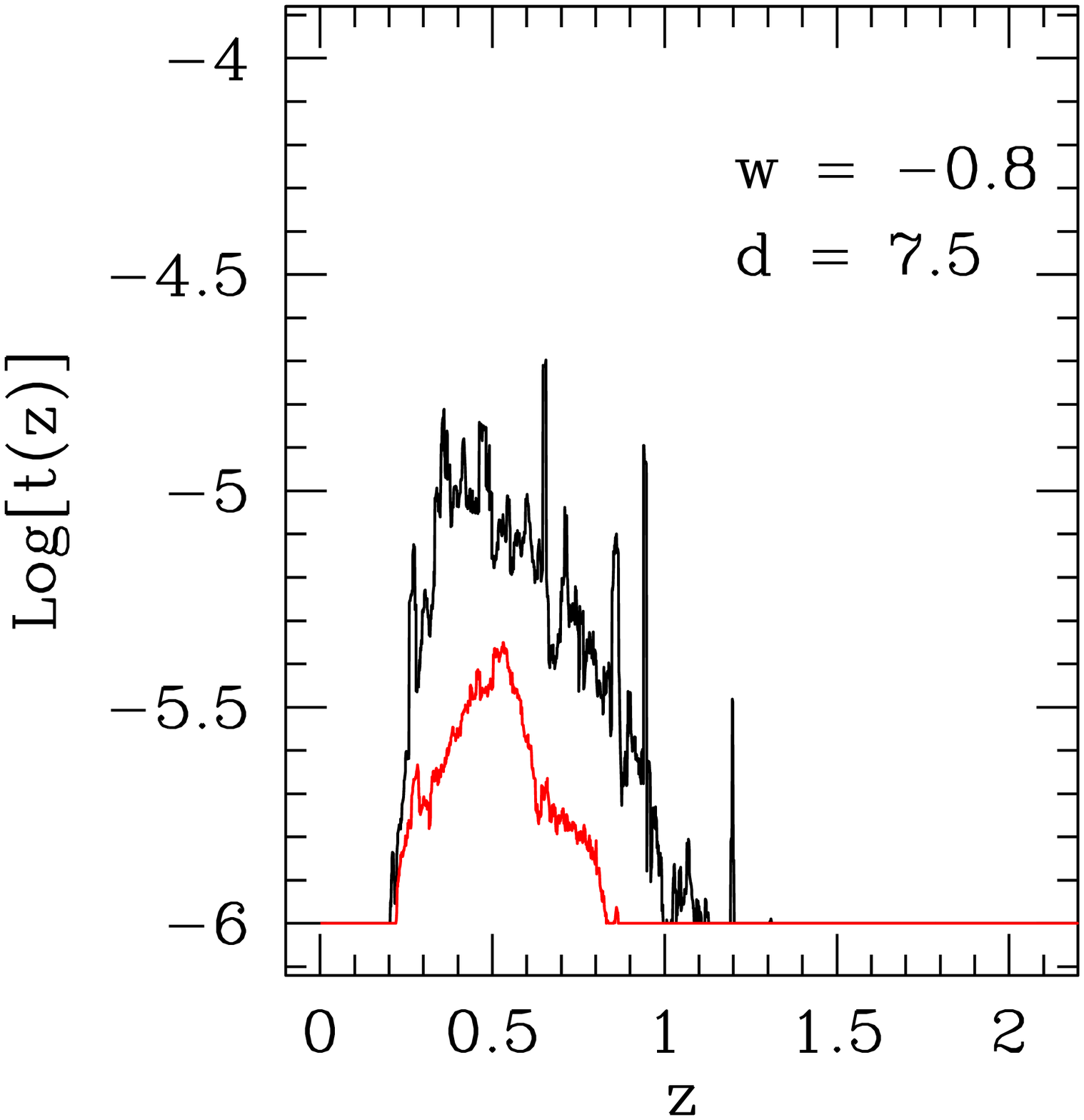}}
  \centerline
   {\includegraphics[width=0.3\hsize]{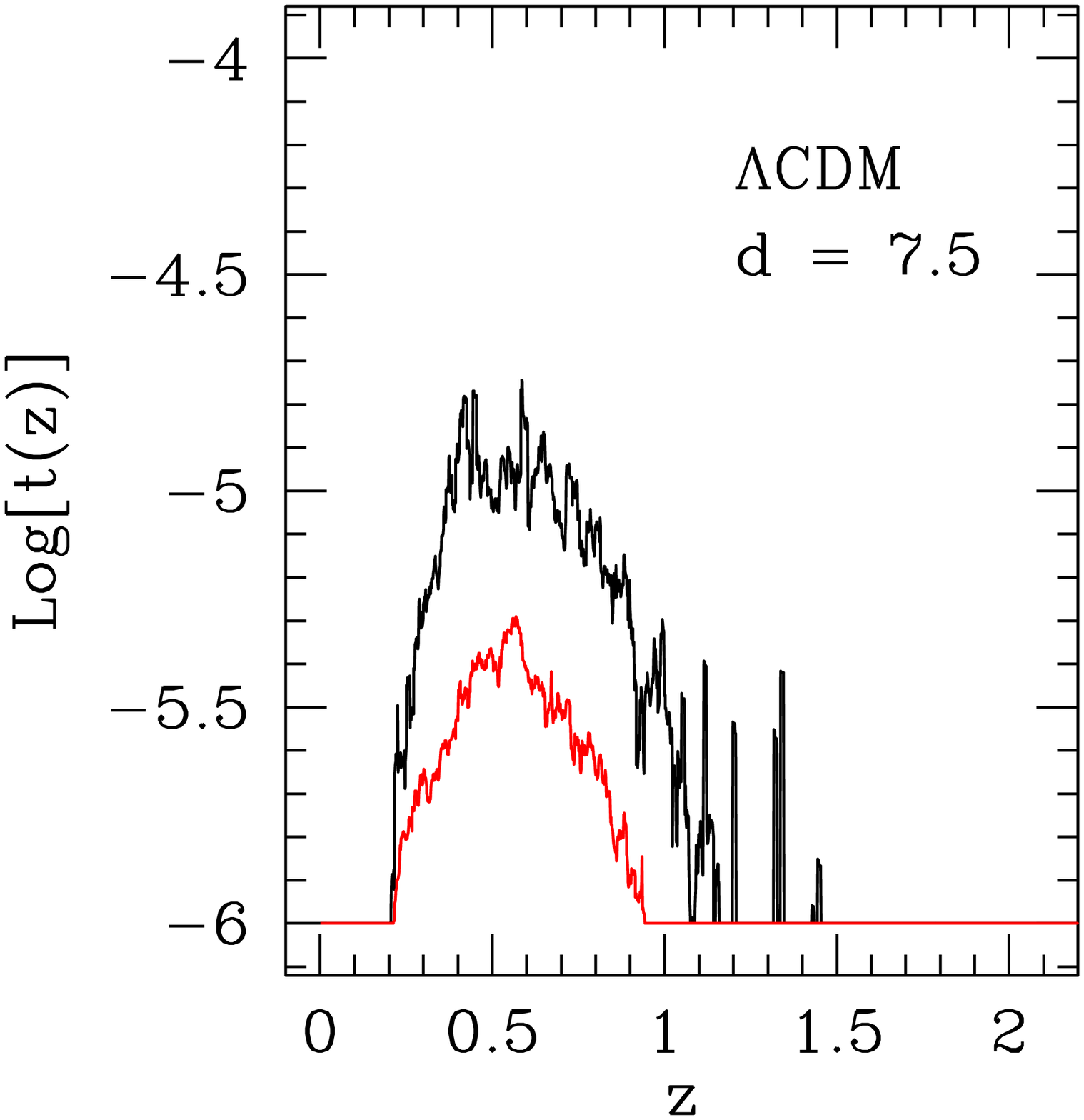}
    \includegraphics[width=0.3\hsize]{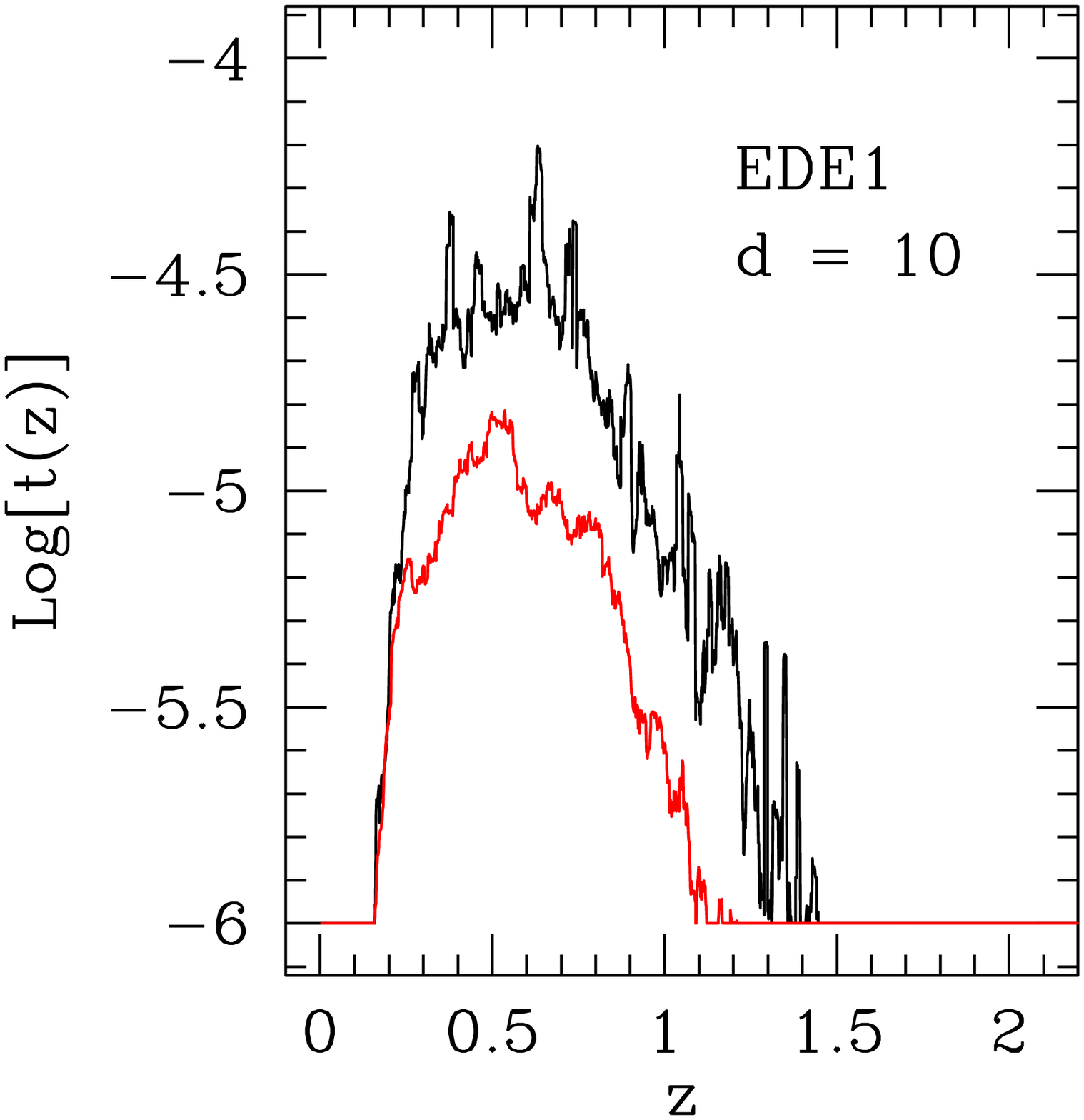}}
\caption{The logarithm of the optical depth per unit redshift for arcs
  with length-to-width ratio exceeding $d=7.5$ for each of the
  cosmological models studied here (top and bottom-left panels), and
  for $d=10$ in the model EDE1 (bottom-right panel). Black curves show
  optical depths obtained including halo mergers with sub-halos, while
  red curves are obtained ignoring the effect of halo interactions.}
\label{fig:boost}
\end{figure*}

This can again be understood in terms of the different dynamics of
structure formation. Keeping the mass of the sub-halo fixed, we expect
more halos of such mass to be available at high redshift with which
the main halo can merge, because structure formation begins earlier in
early dark-energy models. On the other hand, structure growth begins
later in models with a constant equation-of-state parameter and
proceeds more rapidly. Thus, at a sufficiently low redshift, the
abundance of such halos equals that in early dark-energy models,
giving rise to an almost identical merger rate.

It is worth emphasising here that the differences shown between the
different cosmological models are also due, in part or mainly, to the
different normalisation $\sigma_8$ of the power spectrum, which is
chosen to make the models agree with the CMB observations.

Recalling that the source-redshift distribution peaks at redshift
$\sim1.2$, we expect the different merger rates to have little
influence on the optical depth. On the other hand, since the optical
depth is essentially an average of the cross section of different
halos weighted by their relative abundances, we expect the difference
in the mass function to severely affect the strong-lensing
statistics. In early dark-energy models, the optical depth per unit
redshift should exceed those in the $\Lambda$CDM model and the model
with a constant equation-of-state parameter of $w=-0.8$.

We show in the next section how well this expectation is satisfied.

\section{Results}

We discuss now the expected behaviour of the optical depth per unit
redshift (see definition in Sect.~4) in the different dark energy
cosmologies considered in this work. The occurrence of gravitational
arcs is highly sensitive to the abundance and internal structure of
galaxy clusters, which in turn depends on the linear and non-linear
evolution of density fluctuations. We thus expect that the presence
and behaviour of dark energy can affect it.

We show in Fig.~\ref{fig:boost} the optical depth per unit redshift
for arcs with a length-to-width ratio exceeding $d=7.5$, obtained for
each of our four cosmological models. For the model EDE1, we also show
the result for $d=10$. The optical depths accounting for and ignoring
halo mergers are compared. As expected, the lensing efficiency
vanishes near the observer and approaching the source redshift because
of the geometrical drop in lensing efficiency.

Cluster mergers increase the optical depth per unit redshift, and thus
also the total optical depth, factors up to 2 or 3 in all dark-energy
models. The enhancement due to mergers appears more uniform than
obtained by \cite{FE06.1}. This is due to the more than one order of
magnitude larger sample used here and to the much higher time
resolution adopted (up to $10^{-2}$ in redshift).

Quite obviously, increasing the length-to-width threshold decreases
the lensing efficiency, but the features due to merger processes
remain qualitatively the same.

\begin{figure}[t!]
  \includegraphics[width=\hsize]{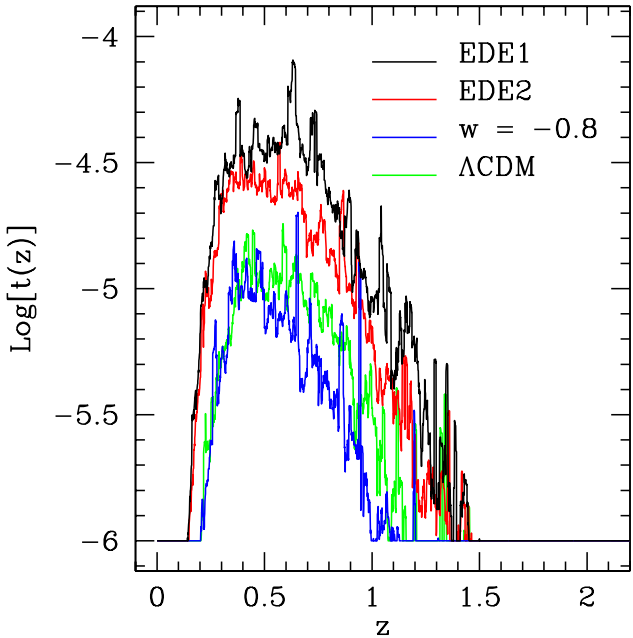}
\caption{The logarithm of the optical depth per unit redshift for arcs
  with length-to-width ratio exceeding $d=7.5$ obtained for the four
  dark-energy models considered here. The lensing efficiency shown
  here takes transient boosts by cluster mergers into account.}
\label{fig:comparison}
\end{figure}

\begin{figure}[t!]
  \includegraphics[width=\hsize]{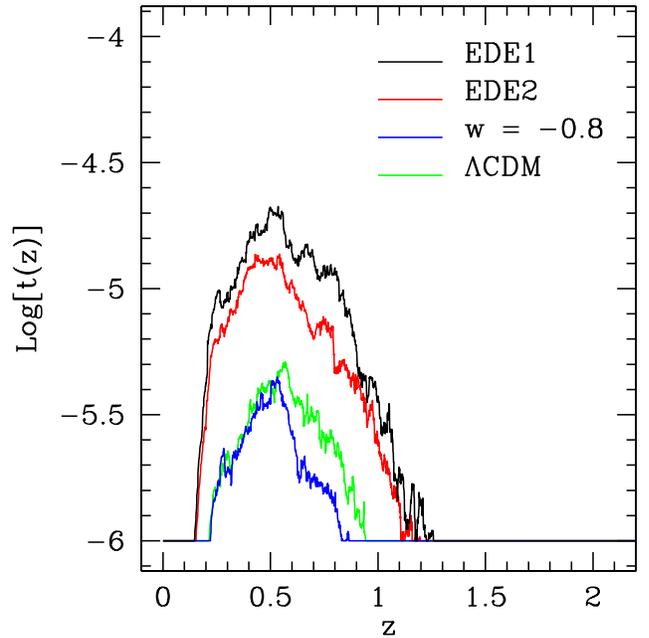}
\caption{Similar to Fig.~\ref{fig:comparison}, but ignoring the effect
  of cluster mergers on the lensing efficiency.}
\label{fig:comparisonWo}
\end{figure}

The main result is that mergers enhance the lensing efficiency by
about the same amount for each model because the merger rate is almost
the same in the redshift range relevant for strong cluster
lensing. However, note that the absolute value of the optical depth
per unit redshift is higher in early dark-energy models, which is
better seen in Figs.~\ref{fig:comparison} and
\ref{fig:comparisonWo}. There, we compare the optical depth per unit
redshift for arcs with length-to-width ratios exceeding $d=7.5$ in the
four cosmologies, accounting for and ignoring cluster mergers,
respectively.

This effect was also expected because of the difference in the
abundance of halos of a given mass in various cosmological
models. These figures show that, both with and without the effect of
halo mergers, the lensing optical depth per unit redshift is higher by
factors up to $\sim3$ in early dark-energy models compared to the
other models. At redshifts above $\sim0.5$, the lensing efficiency for
the model with a constant $w=-0.8$ is slightly smaller than in the
$\Lambda$CDM model because the abundance of halos is also slightly
smaller (see Fig.~\ref{fig:mf}). A similar difference appears between
the EDE1 and EDE2 models. This is due to the fact that in the first
the normalisation of the power spectrum is higher than in the second,
causing a higher abundance of clusters.

An effect that we also recognise in these plots is that in cosmologies
with early dark energy, the optical depth per unit redshift rises and
reaches a significant level already at relatively high redshift, while
it is still negligible in a $\Lambda$CDM model. As discussed before,
the models alternative to $\Lambda$CDM that we have studied here have
a larger fraction of structures at high redshift, causing this earlier
and larger contribution to the strong-lensing efficiency.

\begin{figure*}[ht!]
  \centerline
   {\includegraphics[width=0.3\hsize]{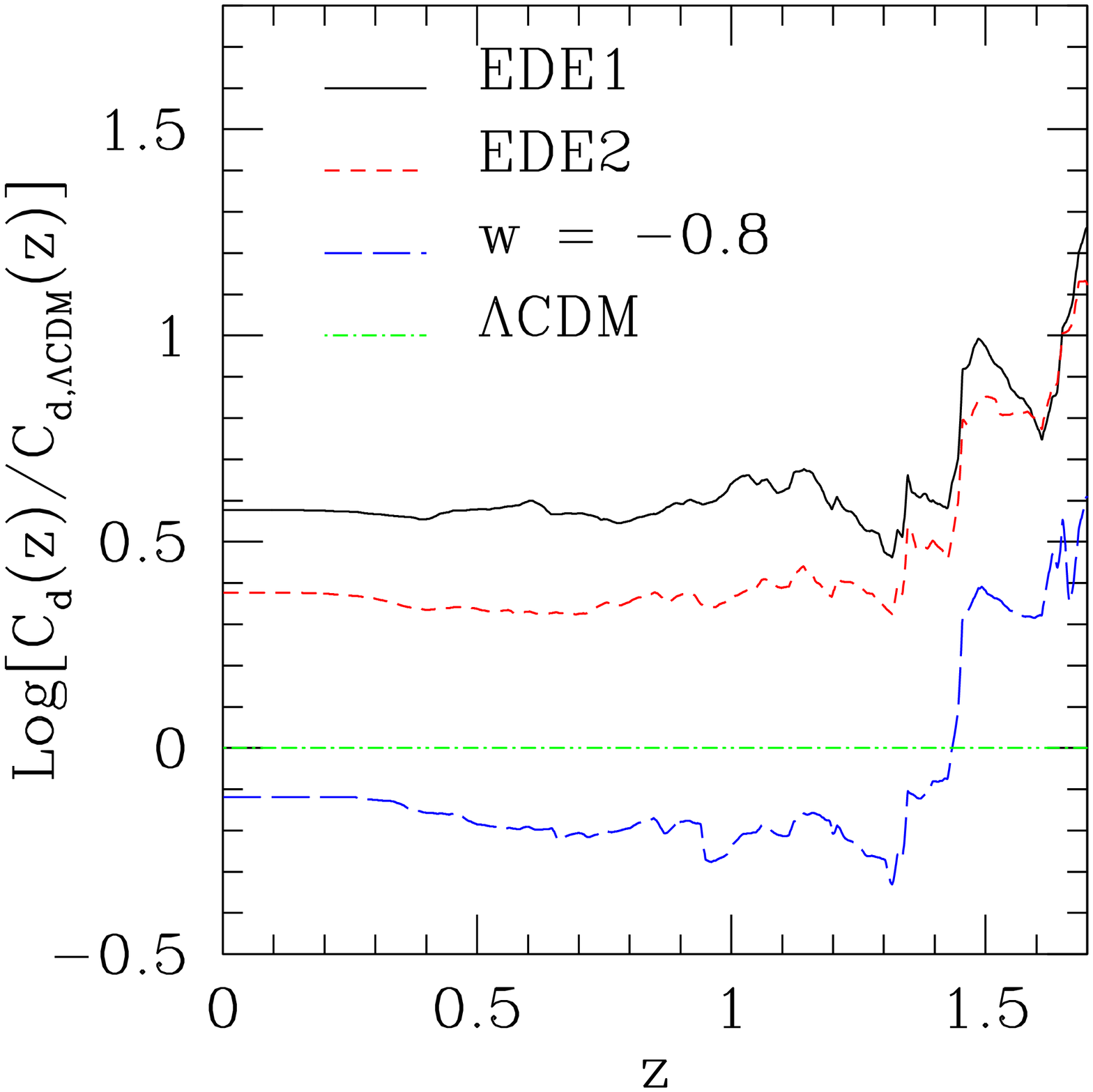}
    \includegraphics[width=0.3\hsize]{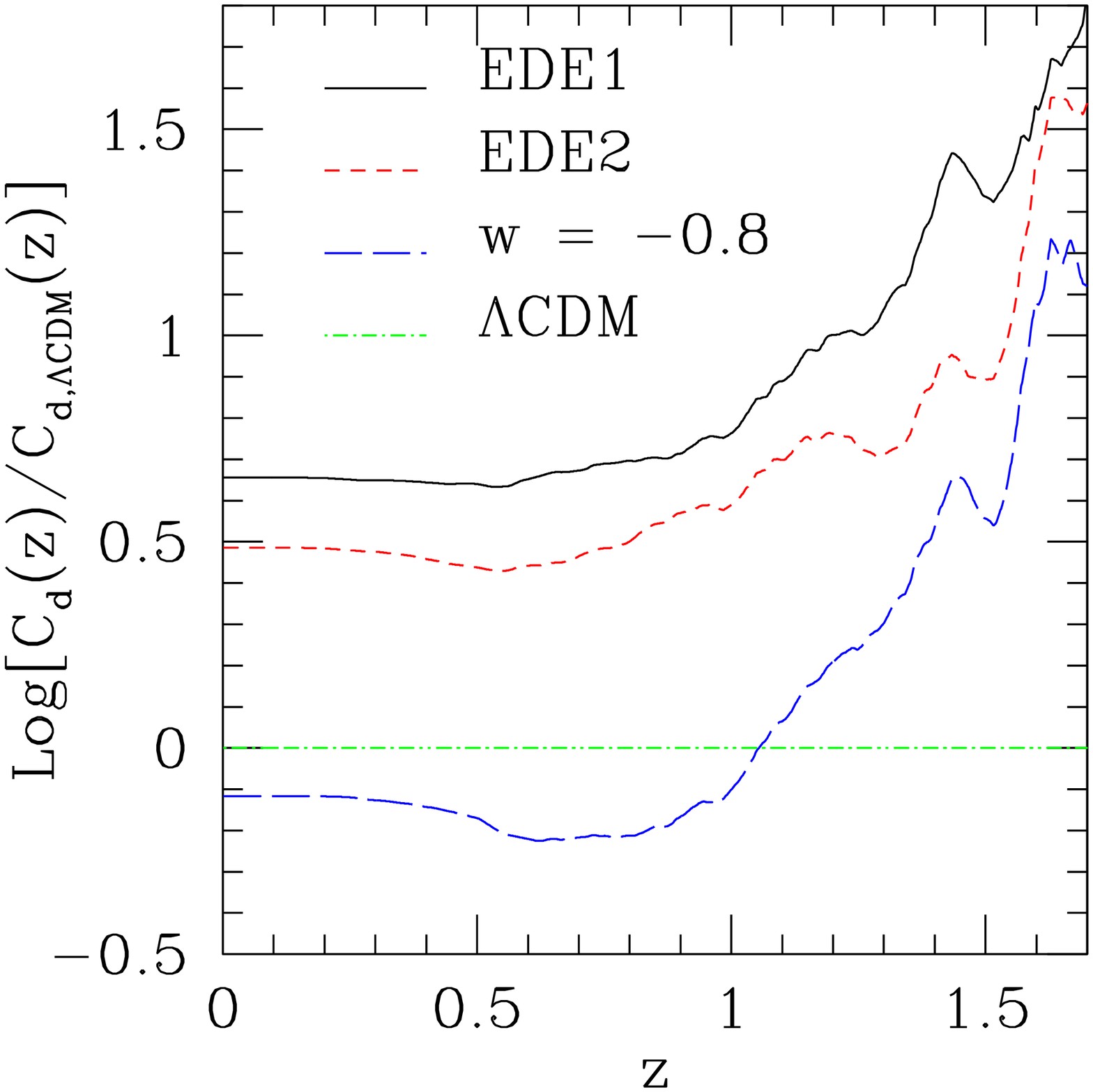}}
  \centerline
   {\includegraphics[width=0.3\hsize]{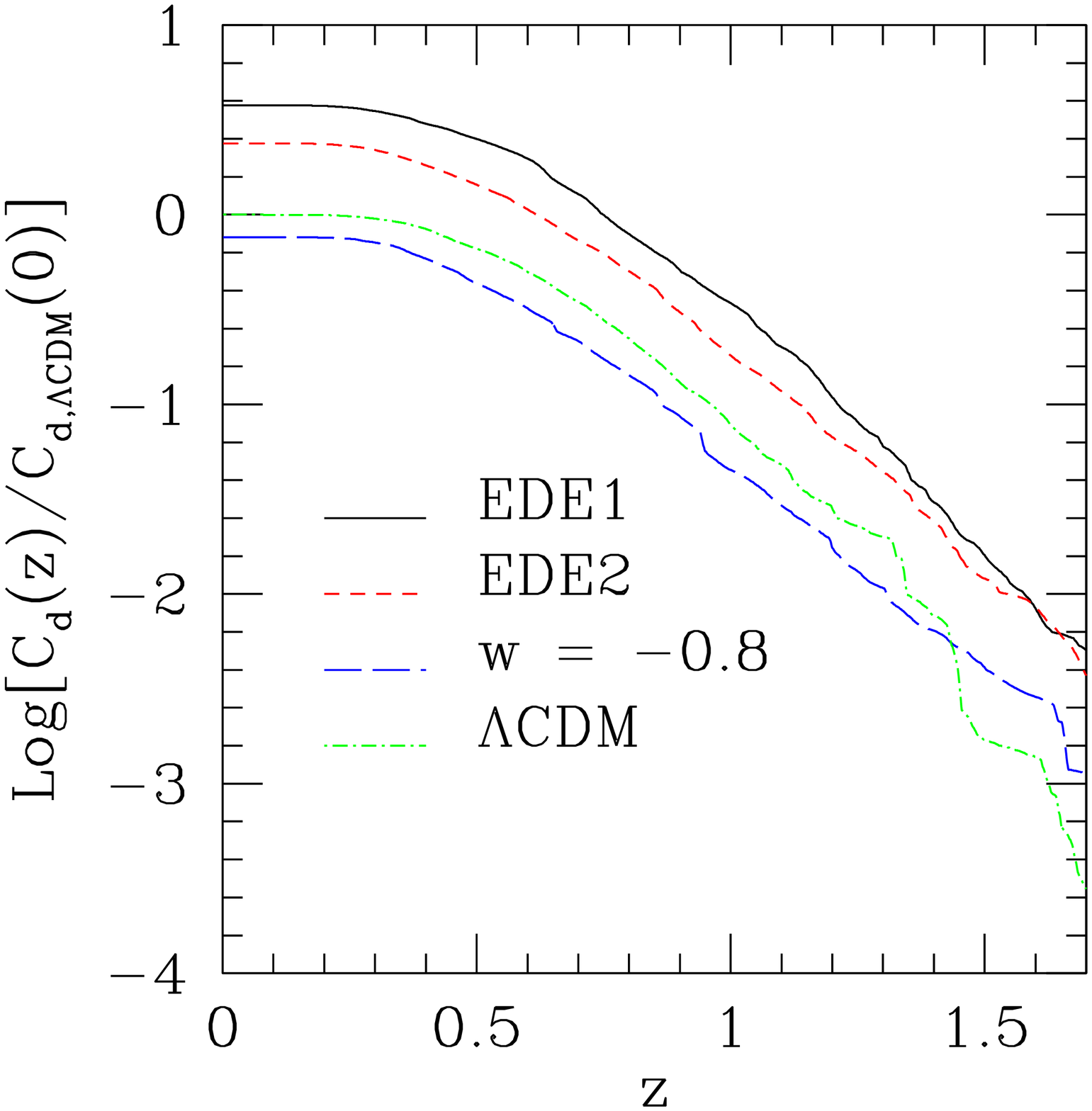}
    \includegraphics[width=0.3\hsize]{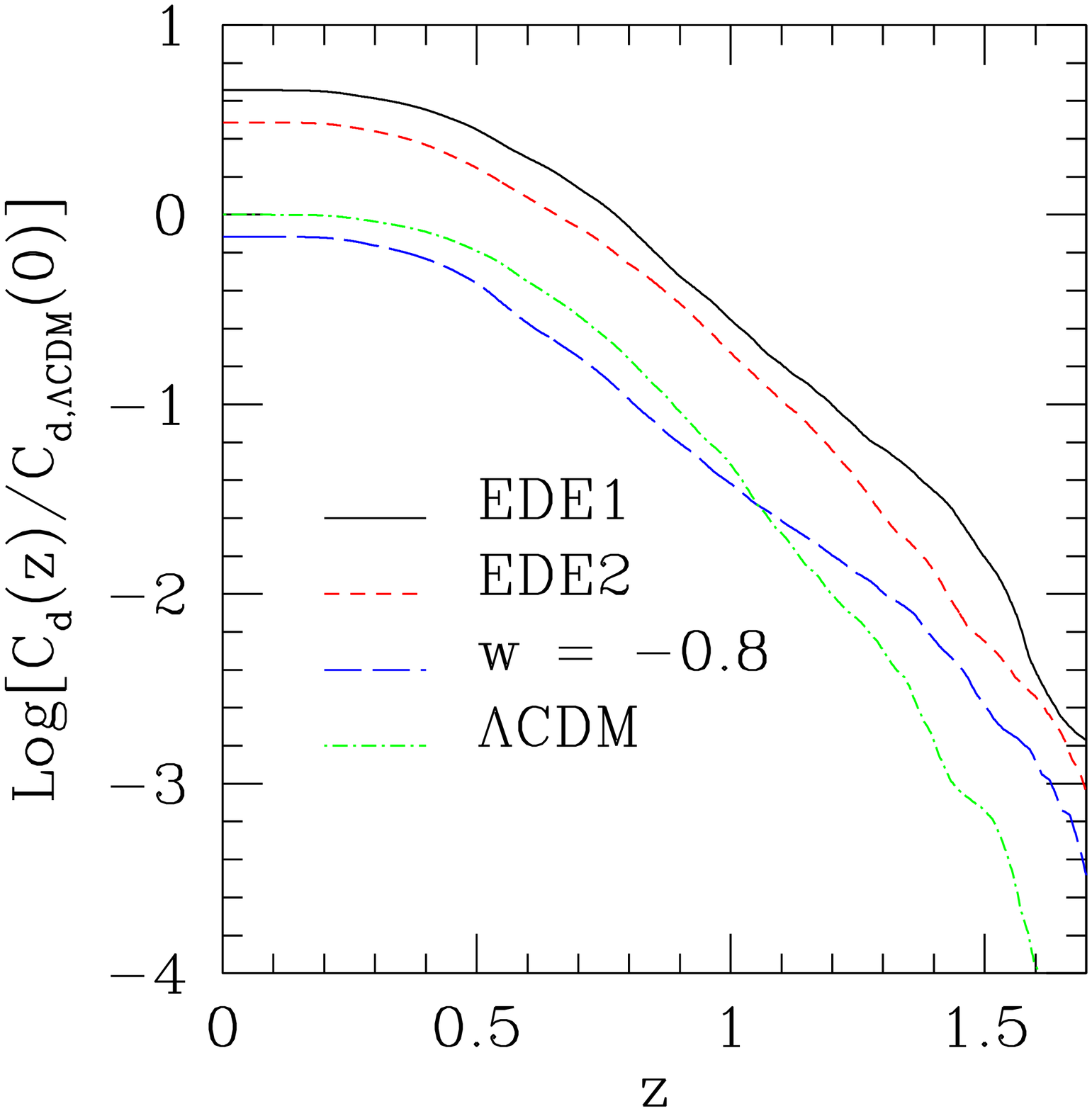}}
\caption{\emph{Top panels}. Logarithm of the cumulative optical depth
  per unit redshift, normalised to its value in a $\Lambda$CDM
  universe. Curves representing the four cosmologies studied in this
  work are shown, as labelled in the plot. \emph{Bottom
  panels}. Logarithm of the cumulative optical depth per unit
  redshift, normalised to its value at present in a $\Lambda$CDM
  universe. Left and right panels show results including and ignoring
  cluster mergers, respectively.\label{fig:cumulative}}
\end{figure*}

Further detail on this aspect is provided by
Fig.~\ref{fig:cumulative}. In its top panels, it shows the cumulative
optical depth per unit redshift, which we can write as
\begin{equation}
  C_\mathrm{d}(z)=\int_z^{z_\mathrm{max}}t_\mathrm{d}(z')\d z'\;,
\end{equation}
normalised to the $\Lambda$CDM case. By its increase towards high
redshift, it emphasises directly how the lensing efficiency drops
already at lower redshift in a $\Lambda$CDM Universe with respect to
the (early) dark-energy cosmologies. The bottom panels show the
cumulative optical depth per unit redshift normalised to the
\emph{present} value in the $\Lambda$CDM model,
$C_{\mathrm{d},\Lambda\mathrm{CDM}}(0)=
\bar\tau_{\mathrm{d},\Lambda\mathrm{CDM}}$. This illustrates the same
effect in a different way. For instance, we see that the cumulative
optical depth per unit redshift in the EDE1 model reaches the same
value $C_{\mathrm{d},\Lambda \mathrm{CDM}}(0)$ already at
$z\approx0.8$ that $\Lambda$CDM reaches today. Conversely, the
cumulative optical depth per unit redshift in the $\Lambda$CDM case
has already dropped by an order of magnitude by
$z\approx0.8$. Similarly, the EDE2 model reaches the total optical
depth of the $\Lambda$CDM model at $z\approx0.7$.

In agreement with our earlier discussion, we note that this specific
evolution does not depend on whether we take dynamical processes into
account or not. The enhanced lensing efficiency in the high redshift
tail may have stimulating consequences, as we shall discuss later.

The large spikes shown in Fig.~\ref{fig:comparison} are obviously due
to the variation of the lensing efficiency of galaxy clusters during
mergers. Very small spikes appear also in Fig.~\ref{fig:comparisonWo},
where dynamical processes are not taken into account. There, they
originate from numerical effects, in particular to the fact that our
time resolution is very high and the number of haloes is limited.
Indeed, the spikes become larger well above redshift unity, where the
number of contributing haloes is reduced (remember that each halo is
characterised by a different source redshift, drawn from a
distribution which peaks around $z\approx1.2$).

\section{Summary and discussion}

We have analysed the incidence of pronounced (long and thin) arcs in
galaxy clusters in four dark-energy models. In particular, we
considered two early dark-energy cosmologies in which the density
parameter in dark energy at high redshift remains small and positive.
We compared them to a model with constant equation-of-state parameter
$w_\mathrm{de}=-0.8$ and a $\Lambda$CDM model for which
$w_\mathrm{de}=-1$.

For each cosmological model, we used Monte-Carlo techniques to build
up merger trees for a set of $\mathcal{N}=500$ cluster-sized dark
matter halos. By modelling each halo by an NFW density profile with
elliptically distorted lensing potential and suitably accounting for
cluster interactions during mergers, we calculated the optical depth
per unit redshift both accounting for and ignoring cluster mergers. To
this end, we also considered a realistic distribution for the source
redshift.

We find that, in agreement with the results of \cite{FE06.1}, cluster
mergers enhance the occurrence of arcs by a factor between 2 and 3.
This occurs in all cosmological models we analysed, and the relative
increase is approximately the same, because the cluster merger rates
in the redshift ranges relevant for strong lensing (below $z\sim1$)
are almost identical (see the discussion in Sect.~6).

However, a potentially more important result is that the optical depth
per unit redshift is larger by a factor of $\sim3$ in early
dark-energy models compared to the models with cosmological constant
or with a constant equation of state parameter $w_\mathrm{de}=-0.8$,
while the differences between the latter two are close to
negligible. There is also a significant difference between the two
early dark-energy models due to the fact that the model EDE1 has a
higher normalisation parameter $\sigma_8$ than EDE2 in order to agree
with the CMB observations.  (cf.~Table \ref{tab:cos}). Thus, halos
form earlier in model EDE1. This is also demonstrated by
Figs.~\ref{fig:mf} and \ref{fig:mr}. Moreover, the lensing efficiency
drops already at a lower redshift in a $\Lambda$CDM Universe than in
the different dark-energy models. The optical depth per unit redshift
has a significant high-redshift tail in early dark-energy cosmologies
while it is negligible otherwise.

A main consequence of these results is that they indicate an
appreciable difference in the incidence of long and thin gravitational
arcs between the $\Lambda$CDM model and models with early dark
energy. Therefore, arc statistics may provide an interesting way to
investigate into the reliability of these models, although the precise
contribution of $\bar{\Omega}_\mathrm{de,sf}$ will probably be better
constrained using cluster counts in the $X$-ray or Sunyaev-Zel'dovich
regimes, which suffer from lower systematics.

The presence of early dark-energy, combined with the transient boosts
due to cluster mergers could help resolve the discrepancy between the
predicted and observed abundances of gravitational arcs. Since
\cite{BA98.2} first pointed out this problem for a $\Lambda$CDM
universe, much discussion developed around this fact
\citep{ME00.1,ME03.1, ME03.2,WA03.1,LI05.1}. At present, it seems that
neither the internal structure of the lensing halos nor the redshift
distribution of the sources can reconcile theory and observations. It
has been shown here that the effects of early dark energy on structure
growth interestingly point into the right direction. Similar
conclusions were drawn also by \cite{ME05.1}, where the lensing
efficiency of numerically simulated dark matter haloes in different
dark energy cosmologies were analysed. Here the haloes are modelled in
an analytical way, allowing a much higher mass and time
resolution. Moreover, the dark energy models studied there were
derived from SUGRA and Ratra-Peebles \citep{PE02.2} potentials,
without an early component. In many aspects, our work is thus
complementary to that of \cite{ME05.1}.

Finally, the fact that the lensing efficiency in early dark-energy
models is much higher at high redshift than in the $\Lambda$CDM case
can be related to the recent unexpected discovery of high incidence of
giant arcs in high redshift clusters \citep{GL03.1,ZA03.1}. Future
searches for strong lensing in distant galaxy clusters may be
promising to distinguish between cosmological models other than the
standard $\Lambda$CDM, or at least to gain a deeper understanding of
the role of early dark energy.
 
\section*{Acknowledgements}

We are grateful to M.~Meneghetti and L.~Moscardini for comments on the
manuscript and useful discussions. This work was supported by the
Collaborative Research Centre SFB 439 of the \emph{Deutsche
Forschungsgemeinschaft}, and the German Academic Exchange Service
(DAAD) under the \emph{Vigoni} programme. We wish to thank the anonymous
referee for useful remarks that allowed us to improve the presentation of the
work.

\bibliographystyle{aa}
\bibliography{./master}

\end{document}